\documentclass[groupedaddress,
 reprint,
 amsmath,amssymb,
 aps,
 prd
]{revtex4-1}

\usepackage{graphicx}
\usepackage{dcolumn}
\usepackage{bm}
\usepackage[hyperindex,breaklinks]{hyperref}

\usepackage{comment}
\usepackage{color}

\begin{document}

\title{Core-halo mass relation of ultralight axion dark matter from merger history}

\author{Xiaolong Du}
\email{xiaolong@astro.physik.uni-goettingen.de}
\author{Christoph Behrens}
\author{Jens C. Niemeyer}
\author{Bodo Schwabe}
\affiliation{%
Institut f\"ur Astrophysik, Georg-August-Universit\"at G\"ottingen, Friedrich-Hund-Platz 1, D-37075 G\"ottingen, Germany
}%

\date{\today}

\begin{abstract}
In the context of structure formation with ultralight axion dark
matter, we offer an alternative explanation for the mass relation of
solitonic cores and their host halos observed in numerical simulations. Our argument is 
based entirely on the mass gain that occurs during major mergers of binary
cores and largely independent of the initial core-halo mass relation assigned to hosts that have just collapsed.
We find a relation between the halo mass $M_h$ and corresponding core mass $M_c$, $M_c\propto M_h^{2\beta-1}$, where $(1-\beta)$ is the core mass loss fraction.
Following the evolution of core
masses in stochastic merger trees, we find empirical evidence
for our model.
Our results are useful for statistically modeling the effects of
dark matter cores on the properties of galaxies and their
substructures in axion dark matter cosmologies.
\end{abstract}


\maketitle

\section{Introduction}\label{sec:intro}

Ultralight scalar fields can be a viable candidate for dark matter if they are in a very cold state (e.g.,\cite{Turner:1983he,Press:1989id,Sin:1992bg,Sahni:1999qe,Hu:2000ke,Goodman:2000tg,Peebles:2000yy,Amendola:2005ad}).
If consisting of particles of mass $\sim10^{-22}{\rm eV}$
\cite{Marsh2014,Schive2014a,Schive2014b,Bozek:2014uqa,Marsh2015b,Schive2016,Marsh2015,Sarkar2016,Calabrese2016,Gonzales-Morales:2016mkl} these candidates can potentially solve the well-known problems faced by pure cold dark matter (CDM) models on small scales (see \cite{Hui:2016ltb} for a recent review).
Possible constituents are ultralight axions (ULAs)
that are  produced
nonthermally via the misalignment mechanism~\cite{Witten:1984dg,Svrcek:2006yi,Arvanitaki:2009fg}.
If self-interactions can be neglected, this type of dark matter candidate is often referred to as fuzzy dark matter (FDM)~\cite{Sahni:1999qe,Hu:2000ke}.
Unlike CDM which produces cuspy halo profiles, FDM produces flat halo cores~\cite{Schive2014a,Schwabe:2016rze,Veltmaat:2016rxo} on scales smaller than the de Broglie wavelength of particles with the
halo's virial velocity, the so-called quantum Jeans
length~\cite{Hu:2000ke,Woo:2008nn}. Below this scale, quantum effects suppress gravitational collapse.

By performing a Jeans analysis, it is found in~\cite{Schive2014a} that the cored halo profile corresponding to FDM with mass $m_a=0.81\times10^{-22}{\rm eV}$ can well reproduce the radial distribution of stars and their velocity dispersion in the Fornax dwarf spheroidal (dSph) galaxy. Further analysis on multiple stellar subpopulations in the Fornax and Sculptor dSph galaxies is done in~\cite{Marsh2015b} and an upper bond, $m_a<1.1\times10^{-22}{\rm eV}$, on the FDM mass is found by assuming that FDM alone can resolve the cusp-core problem. A similar constraint is found in~\cite{Chen:2016unw} from Jeans analysis of eight classical dSph galaxies. In~\cite{Gonzales-Morales:2016mkl}, it is demonstrated that Jeans analysis may be biased due to uncertainties in the assumed halo profile. Instead, a more stringent unbiased constraint, $m_a<0.4\times10^{-22}{\rm eV}$, is obtained in~\cite{Gonzales-Morales:2016mkl} by analyzing the averaged velocity dispersion of dSph galaxies.

Coherent oscillations of FDM also lead to a sharp suppression of the power spectrum~\cite{Hu:2000ke} and halo formation~\cite{Marsh2014,Schive2016,Du:2016zcv,Corasaniti:2016epp} below the Jeans scale. In turn, this cutoff scale for FDM halos puts a lower bound on the FDM mass since deviations from CDM cannot violate the constraints given by current observations.
Using the cosmic microwave background and galaxy clustering data, \cite{Hlozek:2014lca} find
a lower bound on the FDM mass, $m_a\gtrsim10^{-24}{\rm eV}$. Constraints from UV luminosity functions and reionization are much tighter, e.g.~\cite{Corasaniti:2016epp} find $m_a\gtrsim1.6\times10^{-22}{\rm eV}$ (see also~\cite{Bozek:2014uqa} and \cite{Schive2016}). This lower bound is in tension with the upper bound obtained from dwarf galaxies. Furthermore, the Ly$\alpha$ forest also puts a tight constraint on the FDM mass similar to the case of warm dark matter (WDM)~\cite{Viel:2013apy,Marsh2014,Schneider:2016uqi}. Thus, FDM may also suffer from the \emph{Catch 22} problem~\cite{Maccio:2012qf} like WDM, i.e. either producing too small halo cores or too few low-mass halos. However, as discussed in~\cite{Gonzales-Morales:2016mkl}, to get more consistent constraints we need to consider details of the interplay between FDM and baryonic physics. The baryonic feedback may help reconcile the tension between different observations~\cite{Wetzel:2016wro}.

Simulations of cosmological structure formation~\cite{Schive2014a} and
merging solitonic solutions~\cite{Schive2014b,Schwabe:2016rze} based
on the Schr\"{o}dinger-Poisson (SP) equations
indicate that FDM halos contain distinct cores surrounded by Navarro-Frenk-White-like
profiles~\cite{Navarro:1995iw}. \cite{Schive2014a} find that the mass of these cores, $M_c$, is related to the halo mass,
$M_h$, by a power law relation, $M_c\propto M_h^{1/3}$. They propose
an explanation based on the relation
$M_c=\alpha\left(|E|/M\right)^{1/2}$,
where $E$ is the total energy, $M$ is the total mass, and $\alpha$ is a constant of order unity,
which they motivate heuristically with nonlocal consequences of the
Heisenberg uncertainty relation. Identifying $E$ and $M$ with the energy of the halo $E_h$ and its virial mass $M_h$,
they arrive at the numerically measured core-halo mass relation~\cite{Schive2014b}.

Note that while $M_c \sim |E|^{1/2} M^{-1/2}$ is consistent with the intrinsic scaling
properties of the SP equations (see, e.g., \cite{Guzman2004}), it is
not unique (i.e., it can be multiplied by any scale invariant combination of $|E|$ and $M$). Removing any residual effects of the scaling symmetry by
constructing and analyzing scale invariant quantities, \cite{Schwabe:2016rze} were
unable to reproduce this relation in simulations of solitonic core
mergers. Furthermore, the model of \cite{Schive2014b} does not account for the
combined evolution of $M_c$ and $M_h$ by halo mergers after the
initial collapse of density perturbations which is known to be an
important ingredient in hierarchical structure formation.

Comparing the initial and final masses of merging cores,
\cite{Schwabe:2016rze} find a universal behavior of the core mass loss in mergers that
depends nearly entirely on the mass ratio. 
Implementing
this relation in a semianalytic model (SAM) for galaxy formation,
\cite{Du:2016zcv} studies the effects of the core on the substructure of
Milky way-sized FDM halos.

Here, we present a model for the core mass as a function of halo mass 
which is entirely based on the fractional
core mass loss during major mergers.
No further assumptions about the quantum nature of FDM halos and cores are
necessary. In particular, our model is independent of the dynamics of
halo formation by gravitational collapse and hence insensitive
to the initial core-halo mass relation of newly formed halos. 
We find a simple relation between the core and halo mass whose slope 
is a function of the core mass loss fraction. 
We provide numerical evidence
for this dependence using stochastic merger trees. 

The existence of compact cores in the halo substructure has many
potentially observable direct signatures in, for instance, rotation curves of
dwarf galaxies \cite{Rhee:2003vw,Oman:2015xda}, gravitational lensing \cite{Mao:1997ek,Metcalf:2001ap,Hezaveh2014,Hezaveh2016}, globular
cluster streams in the Milky Way \cite{Odenkirchen:2000zx,Grillmair:2006bd}, or the thickening of the thin galactic disk
\cite{Toth1992,Quinn1993,Navarro:1994zk,Walker:1995ef,Sellwood:2013npa,Benson:2003uc,Kazantzidis:2007hy}. Indirectly, the effects of compact cores on star
formation at high redshifts may be probed by the reionization history
and the high-$z$ galaxy luminosity function. 

In addition to providing a simple explanation for the core-halo mass
relation, it is straightforward to produce realizations of our
stochastic model from modified EPS merger trees \cite{Du:2016zcv}.
Since--as we will show--the core mass is
determined by the individual accretion history, it 
can be modeled more realistically using individual mass accretion
histories that recover not only the mean core-halo mass relation but
also its scatter.

\section{Core-Halo Mass Relation}\label{sec:results:chmr}

Given the mass loss fraction of cores during each merger, we can
calculate the evolution of core mass along the merger history.
Since cores merge and relax to their final state on a dynamical time
scale once they begin to overlap
\cite{Schwabe:2016rze}, we only need to consider isolated binary mergers.

To calculate the evolution of core masses, we first need to know the
initial core masses for halos without progenitors, i.e., those that
form from direct collapse. Their mass is determined
by the cutoff mass in the halo mass function (HMF).
As shown in \cite{Du:2016zcv}, the cutoff mass depends only mildly on
redshift, so the directly collapsed halos have approximately equal masses,
$M_{\rm h,min}$, independent of their collapse redshift. The initial core
masses are therefore also expected to have roughly equal values, $M_{\rm c,min}$. 

Halo mergers change both core and halo masses.
\cite{Schwabe:2016rze} find that only mergers with mass ratio
$\mu<7/3$ yield an increased core mass $M_c = \beta (M_{c1}+M_{c2})$. 
Here, $M_{c1}$ and $M_{c2}$ are the masses of the initial cores and
$(1-\beta)$ is the core mass loss fraction, where $\beta\sim 0.7$
independent of the initial
core masses. We refer to mergers with $\mu<7/3$ as major
mergers. Larger mass ratios (minor mergers) leave the core mass of the more massive
halo unchanged and result in the total disruption of the smaller
halo. Smooth accretion corresponds to accretion with very high mass ratios and is treated in the same way.

In summary, in our model three types of physical interactions can change the core and
halo masses: smooth accretion (see \cite{Du:2016zcv} for more
details), minor mergers, and major mergers. 
The first two increase the mass of the halo but leave the core mass
unchanged. Major mergers increase both halo and core
masses. 

Our model is based on a simplified description of the merging
process. Suppose that 
$N$ halos with halo mass $M_{\rm h,min}$ and core mass $M_{\rm c,min}$
merge to form one halo with mass $M_h$
whose contribution from mergers is $N M_{\rm h,min}$.
Assuming that the mass contributed by smooth accretion is also proportional to $N
M_{\rm h,min}$, we have 
\begin{equation}
M_h=\alpha N M_{\rm h,min}.
\label{M_h}
\end{equation}
If the final halo encounters $N_{\rm major}$ major mergers and $N_{\rm minor}$ minor mergers, then
\begin{equation}
N=N_{\rm major}+N_{\rm minor}+1.
\label{N}
\end{equation}
The more major mergers the final halo encounters, the more minor mergers it also tends to have. Hence, we can assume
\begin{equation}
N_{\rm minor}=b(\beta)N_{\rm major}.
\label{N_minor}
\end{equation}
We will show below that the assumptions Eqs.~(\ref{M_h}) and (\ref{N_minor}) are reasonable.
Given the mass of the final halo, Eqs.~(\ref{M_h}), (\ref{N})
and (\ref{N_minor}) allow us to estimate the number of major mergers it has encountered:
\begin{eqnarray}
N_{\rm major}&=&\frac{1}{1+b(\beta)}\left(\frac{M_h}{\alpha M_{\rm h,min}}-1\right)\nonumber\\
&\approx&\frac{1}{1+b(\beta)}\frac{M_h}{\alpha M_{\rm h,min}}.
\label{N_major}
\end{eqnarray}

Since minor mergers do not change the core mass, we only need to
consider major mergers when estimating the final core mass
$M_c$. Suppose that during every 
major merger, both progenitors have the same core mass, i.e. the
$(N_{\rm major}+1)$ first-formed halos with core mass $M_{c,min}$ merge
pairwise and form $\frac{N_{\rm major}+1}{2}$ halos with core mass
$2\beta M_{\rm c,min}$. This process continues until the formation of the
final halo with mass $M_h$. 
The other $(N-N_{\rm major}-1)$ first-formed halos are 
assumed to be accreted by minor mergers, thus they do not affect the core
mass. 

As explained above, the first-formed halos have nearly identical core
masses, hence the assumption that all major mergers have core mass
ratio $\mu=1$ is reasonable for the first generation of merging events. As halos
continue to merge, we 
overestimate the core mass
because there will be major mergers with
$\mu > 1$ and correspondingly smaller core mass growth.

Finally, after
$\log_2(N_{\rm major}+1)$ generations of major merger events and $N_{\rm minor}$ minor merger events, the final halo has a core mass of
\begin{eqnarray}
M_c&=&(2\beta)^{\log_2 (N_{\rm major}+1)}M_{\rm c,min}\nonumber\\
&=&(N_{\rm major}+1)^{\log_2 (2\beta)}M_{\rm c,min}\nonumber\\
&\approx&(N_{\rm major})^{\log_2 (2\beta)}M_{\rm c,min}.
\label{M_c_N_major}
\end{eqnarray}
Substituting Eq.~(\ref{N_major}) into Eq.~(\ref{M_c_N_major}), we have
\begin{equation}
M_c=\left[\frac{1}{1+b(\beta)}\frac{M_h}{\alpha M_{\rm h,min}}\right]^{\log_2 (2\beta)}M_{\rm c,min}
\equiv A M_h^{\log_2 (2\beta)}.
\label{M_c_M_h}
\end{equation}
Note that although the relation Eq.~(\ref{M_c_M_h}) does not
explicitly depend on redshift, the prefactor $A$ does since
$\alpha$ and $b$ change with redshift. On the contrary, the
exponent of $M_h$ only depends on the core mass loss fraction.
In a binary merger, the core mass of the descendant will not be larger than the sum of core masses of its two progenitors, so $\beta\le1$ (e.g. \cite{Schwabe:2016rze} found $\beta\sim0.7$).
Treating $(2\beta-1)$ as a small number, Eq.~(\ref{M_c_M_h}) yields
\begin{equation}
M_c\propto M_h^{\log_2 (2\beta)} \approx M_h^{(2\beta-1)/\ln2}\approx M_h^{1.44(2\beta-1)}
\label{M_c_M_h_s}
\end{equation}
to leading order.
As discussed above, this relation overestimates the core mass when binary mergers with $\mu>1$ are involved. 
We will account for this effect below when presenting our numerical results.

In order to test the core-halo mass relation given in
Eq.~(\ref{M_c_M_h}), we use the modifications to the SAM code
G{\footnotesize ALACTICUS} \cite{Benson2012,Benson:2012su} for FDM described in \cite{Du:2016zcv}
and build $2000$ merger trees for root halos with
$4\times10^{11}<M_{h}<4\times10^{13} M_{\odot}$. The mass resolution is set to $2\times10^8 M_{\odot}$.
Without loss of generality, we set $m_a=10^{-22}{\rm eV}$. The parameter $\beta$ is set to $0.7$ as found by~\cite{Schwabe:2016rze} unless specified otherwise.

Using G{\footnotesize ALACTICUS}, we first construct the merger history for each root halo by successively drawing branching events backward in time until the halo mass of the progenitors is below the mass resolution. The branching rate is calculated from the extended Press-Schechter formalism~\cite{Press:1973iz,Bond:1990iw,Lacey:1993iv}. Halos which have no progenitors are then evolved forward in time, taking into account different physical effects such as mergers, dynamical friction and tidal stripping. The core mass is traced along the merger history and recalculated
at each merger event.

Equation~(\ref{M_c_M_h})
predicts that while the proportionality factor $A$ may depend on the initial core mass $M_{c,min}$, the exponent is independent of it.

\begin{figure*}
\includegraphics[width=0.32\textwidth]{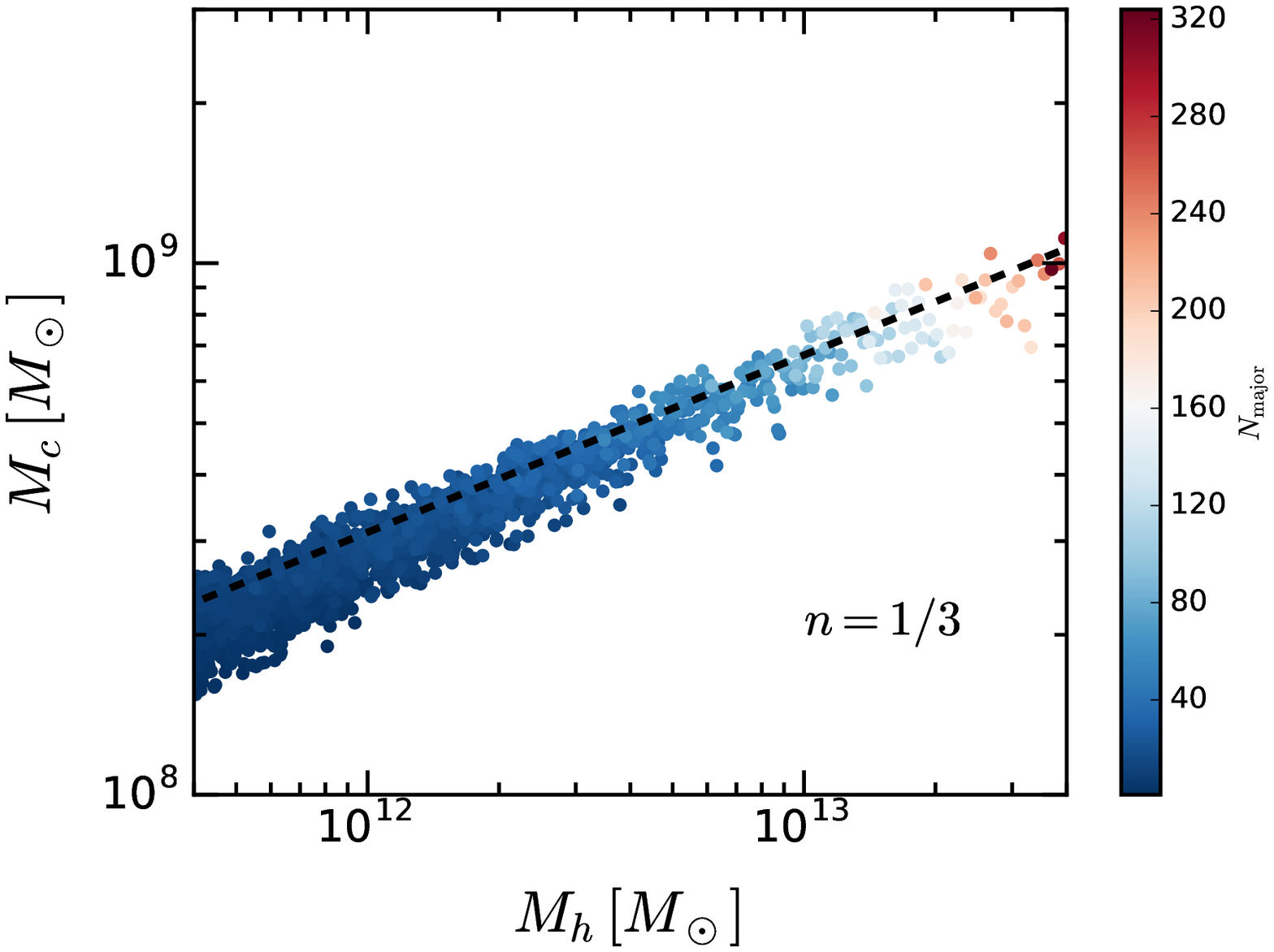}
\includegraphics[width=0.32\textwidth]{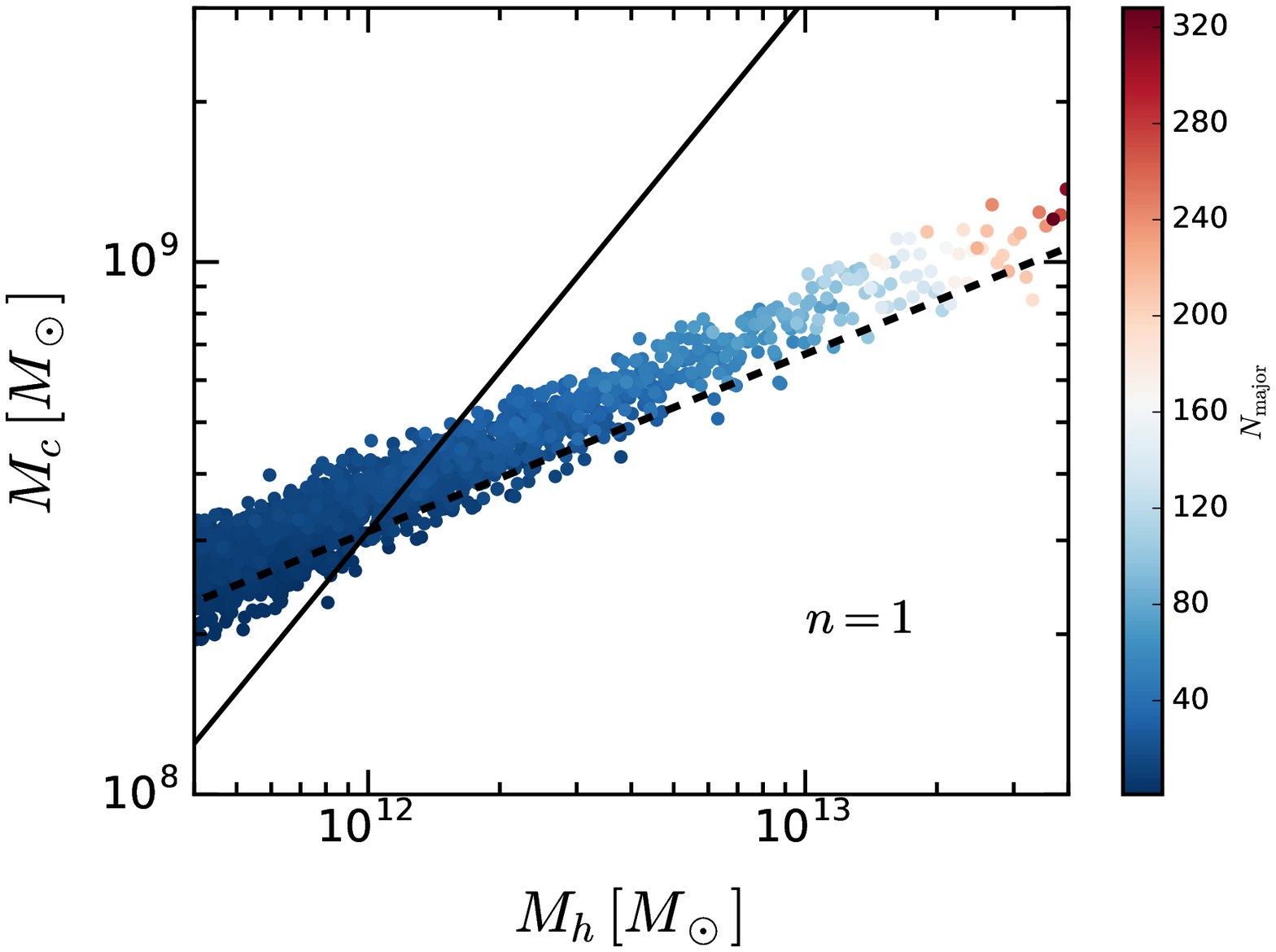}
\includegraphics[width=0.32\textwidth]{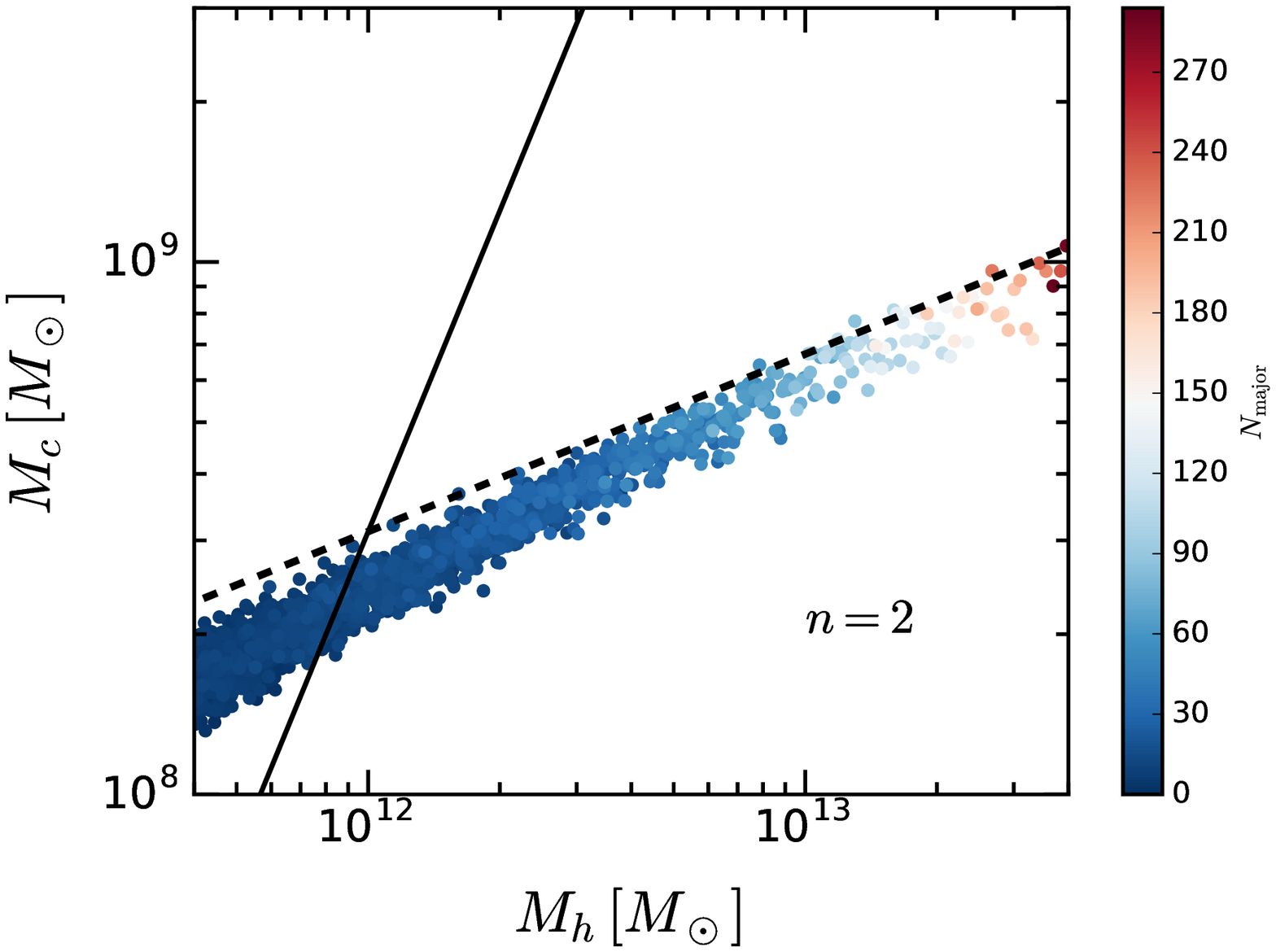}
\caption{The core mass with respect to the halo mass at $z=0$ for
  different initial core-halo mass relation: $n=1/3$ (left), $n=1$
  (center), and $n=2$ (right). The dashed line shows the core-halo
  mass relation from \cite{Schive2014b},  $n=1/3$, at $z=0$. The solid lines show the
  linear and square relations for comparison.}  
\label{fig:Mc_Mh_power}
\end{figure*}

To test the dependence of the core-halo mass relation on initial conditions,
we implemented a power-law initial relation $M_{\rm c,ini}\propto M_{\rm h,ini}^n$ for halos that have no progenitors
in the modified G{\footnotesize ALACTICUS} code. Figure~\ref{fig:Mc_Mh_power} shows the results
for $n=1/3$
, $n=1$
, and $n=2$.
Clearly, the core-halo
mass relation at $z=0$ depends only very weakly on the initial mass distribution. 

In \cite{Schive2014b}, the $1/3$ power-law relation between
the core and halo mass is explained via the uncertainty principle. Although this approach may not be valid for halos which have encountered many mergers, it is applicable for halos that have just collapsed. Since there is no other preferred choice, we will use
$n= 1/3$ to set the initial core mass below. As is shown above, this specific choice does not have a
significant effect on our results.

\begin{figure*}
\includegraphics[width=0.32\textwidth]{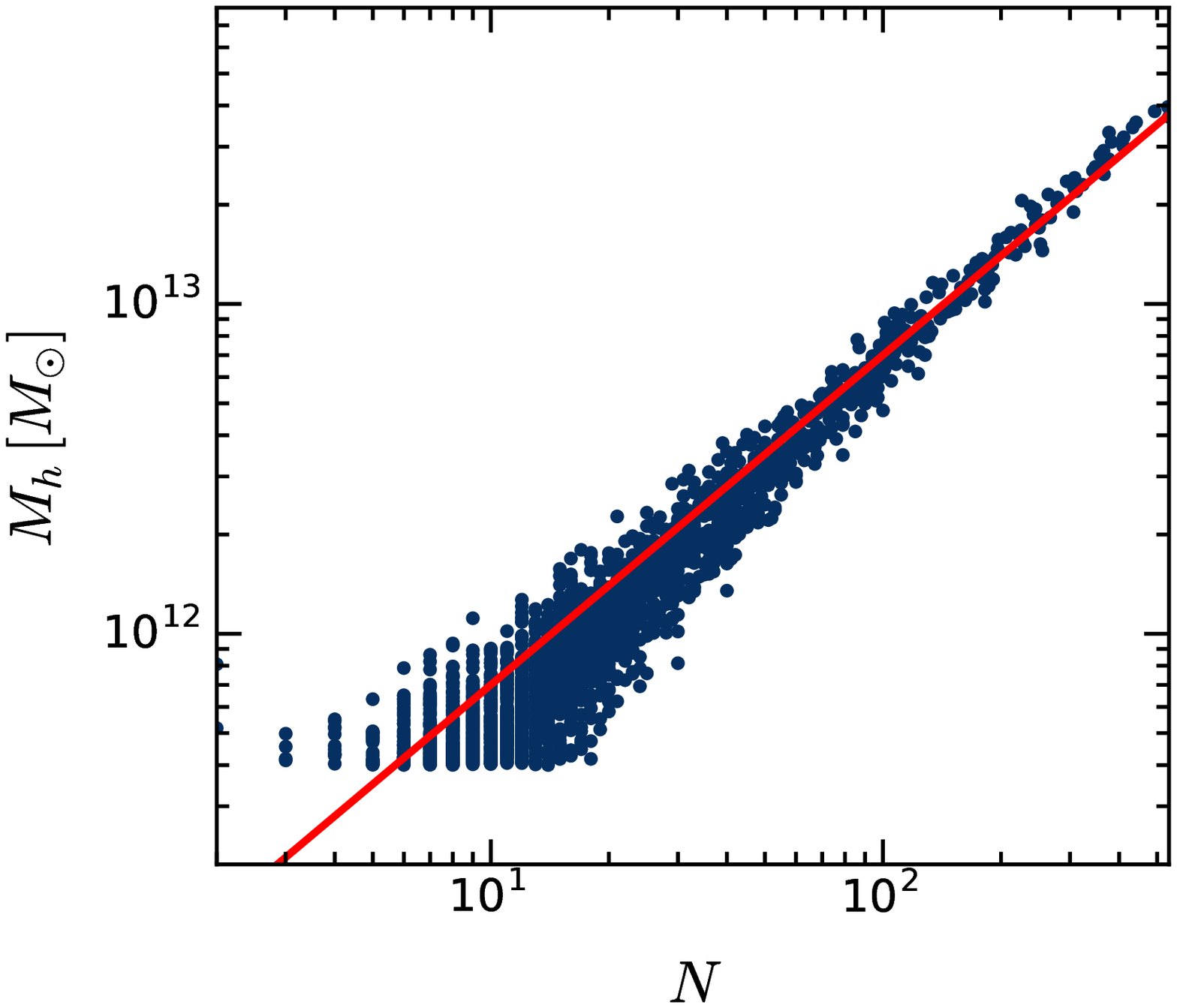}
\includegraphics[width=0.32\textwidth]{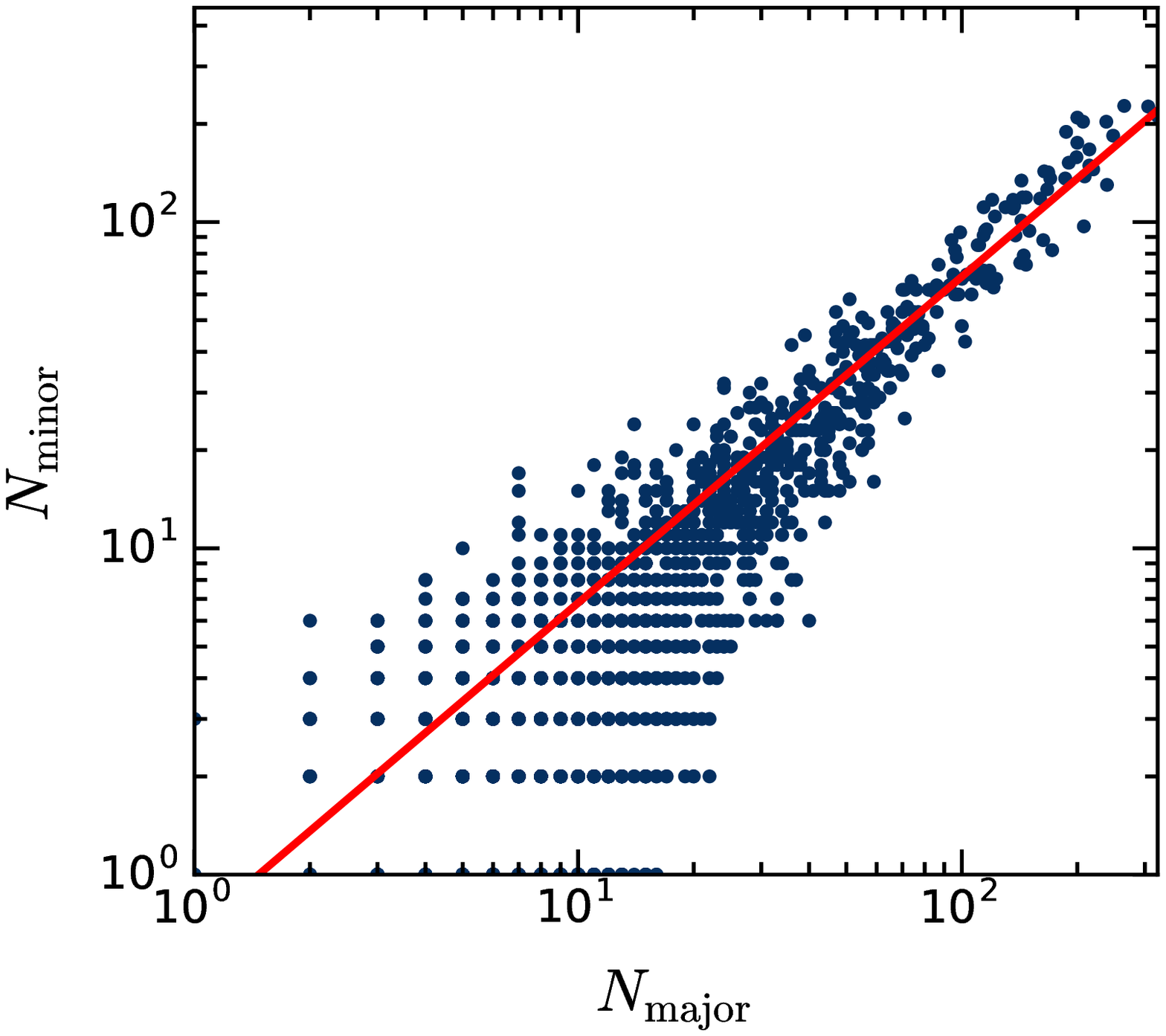}
\includegraphics[width=0.32\textwidth]{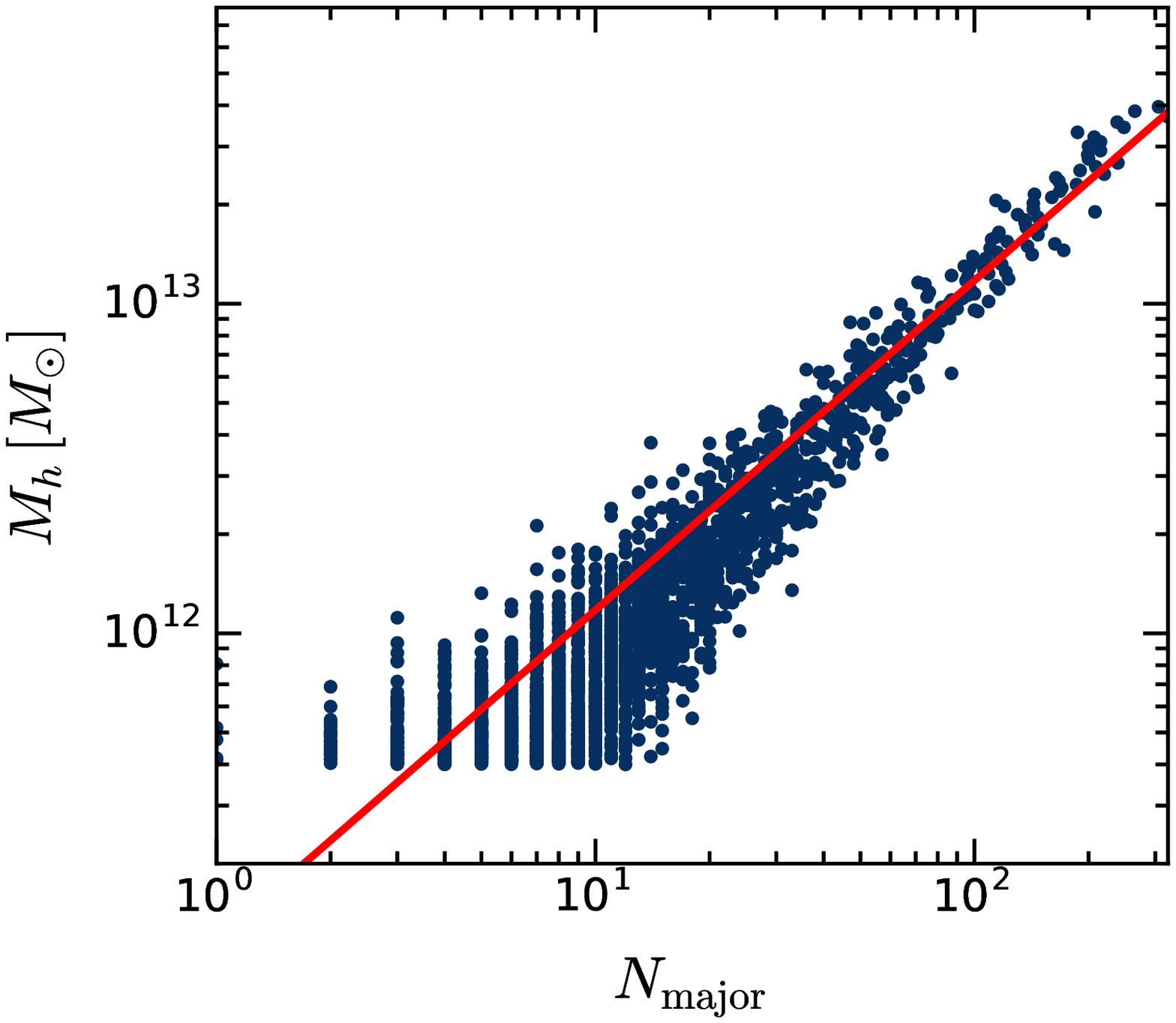}
\caption{Left: the halo mass with respect to the number of
  first-formed halos. The solid line
corresponds to Eq.~(\ref{M_h}).
Center: the number of minor mergers with respect to the number of
major mergers [solid line
given by Eq.~(\ref{N_minor})].
Right: the halo mass with respect to the number of major mergers
[solid line
given by Eq.~(\ref{N_major})].} 
\label{fig:N_Mh}
\end{figure*}

Next, we verify the two assumptions made in deriving 
Eq.~(\ref{M_h}) and Eq.~(\ref{N_minor}).
The left panel of Fig.~\ref{fig:N_Mh} shows the halo mass $M_h$ with
respect to the number of first-formed halos $N$ obtained from merger
trees. Despite large scatter
at small $N$ representing
halos that have only encountered few mergers and are thus more
strongly affected by the uncertainty of
individual events, the 
assumed linear dependence Eq.~(\ref{M_h}) fits well.

The center panel of Fig.~\ref{fig:N_Mh} shows the number of minor
mergers $N_{\rm minor}$ with respect to the number of major mergers
$N_{\rm major}$. Again, at small $N_{\rm major}$ the data points have large
scatter, but in general
the assumption Eq.~(\ref{N_minor}) gives a reasonable fit.
Finally, the right panel of Fig.~\ref{fig:N_Mh} shows the halo mass
$M_h$ with respect to the number of major mergers $N_{\rm major}$. This
plot is a combination of the first two and is just meant to give a
more relevant comparison between Eq.~(\ref{N_major}) inferred from the
two assumptions and the results from merger trees. 
 
To study the impact of the core mass loss fraction, we varied the value of parameter $\beta$ between $0.5$ and $1$.
Correspondingly, we must also modify the definitions of minor and major
merger: if the core mass ratio is larger (smaller) than $\beta/(1-\beta)$, the
merger is defined as minor (major) merger. For $\beta=0.7$, we obtain the former definition. 

Before showing the results from merger trees, we consider two extreme cases.
In the case of $\beta=0.5$, the core mass does not change during any of the three possible interactions.
The final core mass is solely determined by the initial core mass and independent of the final halo mass.
On the contrary, for $\beta=1$, all mergers will be major mergers and
the final core mass is given by $M_c=N
M_{\rm c,min}$. Since the halo mass is also proportional to $N$
[Eq.~(\ref{M_h})], in this case 
the core-halo mass relation is linear.
Expressed in the form $M_c\propto M_h^{\gamma(\beta)}$, we thus have
$\gamma(0.5)=0$ and $\gamma(1)=1$. 
A simple linear parametrization for
$\gamma(\beta)$
is $2\beta-1$ which yields the core-halo mass relation
\begin{equation}
M_c\propto M_h^{2\beta-1}.
\label{M_c_simple}
\end{equation}
Note that it is very similar to Eq.~(\ref{M_c_M_h_s}) 
obtained from the merger history.

\begin{figure*}
\includegraphics[width=0.32\textwidth]{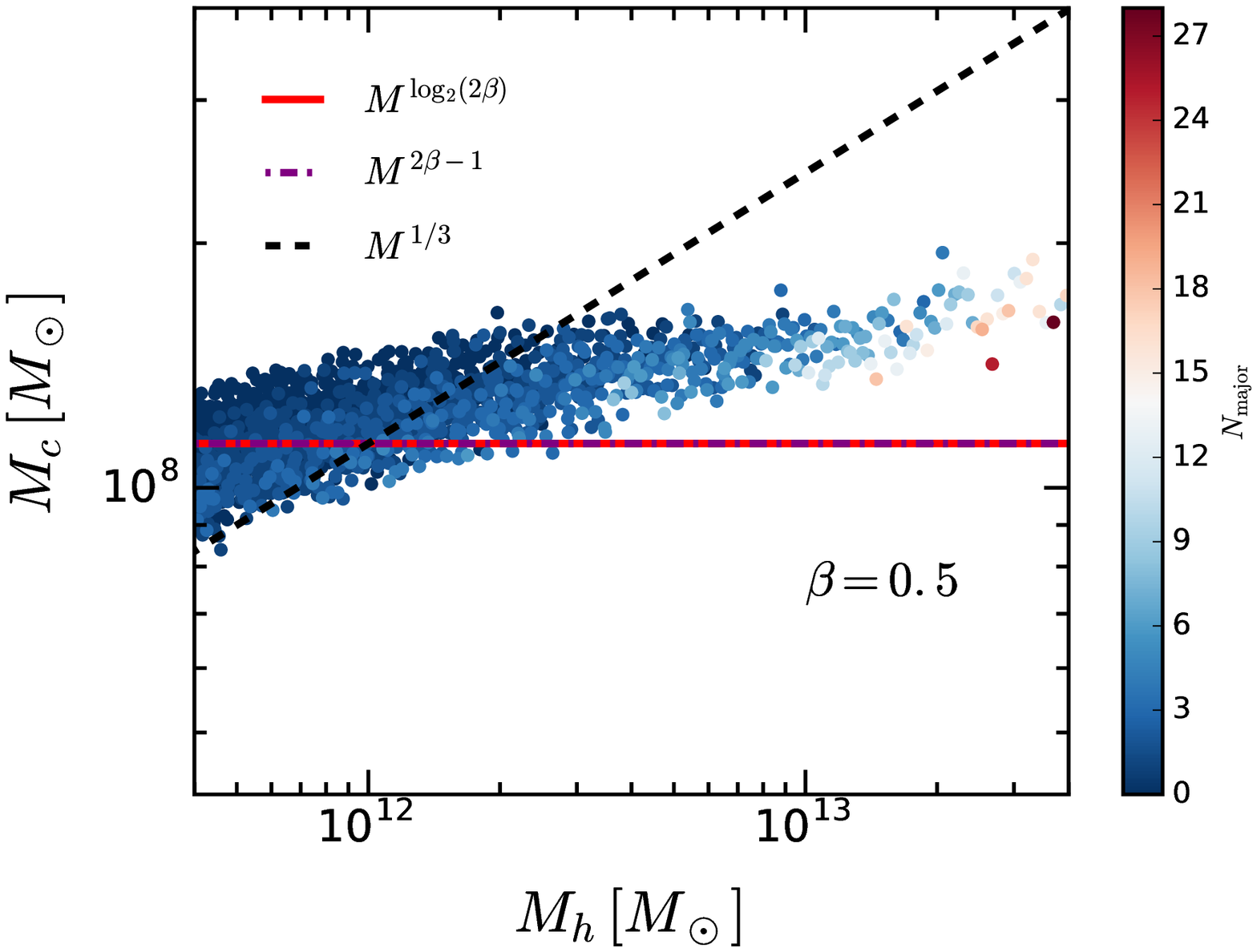}
\includegraphics[width=0.32\textwidth]{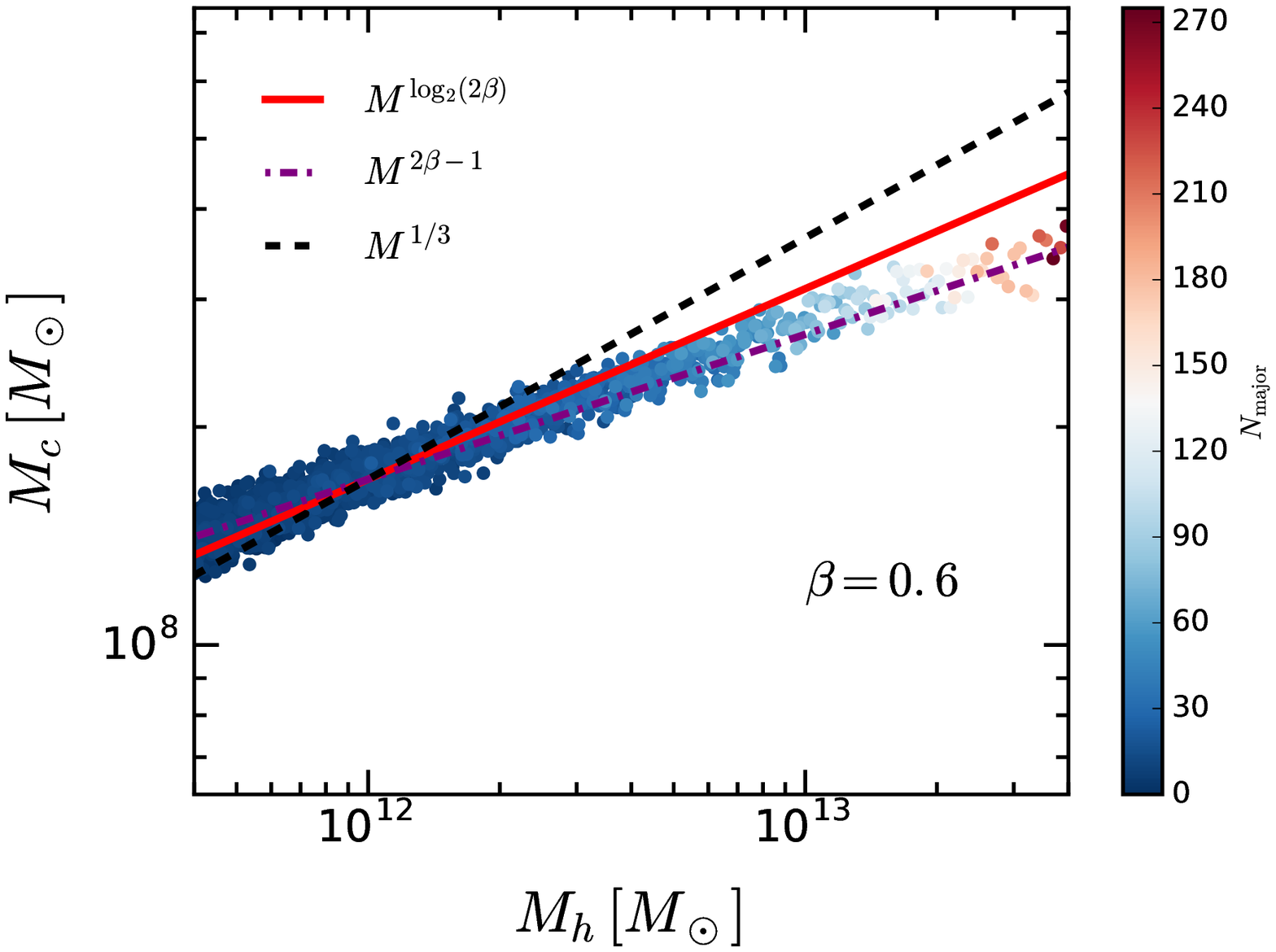}
\includegraphics[width=0.32\textwidth]{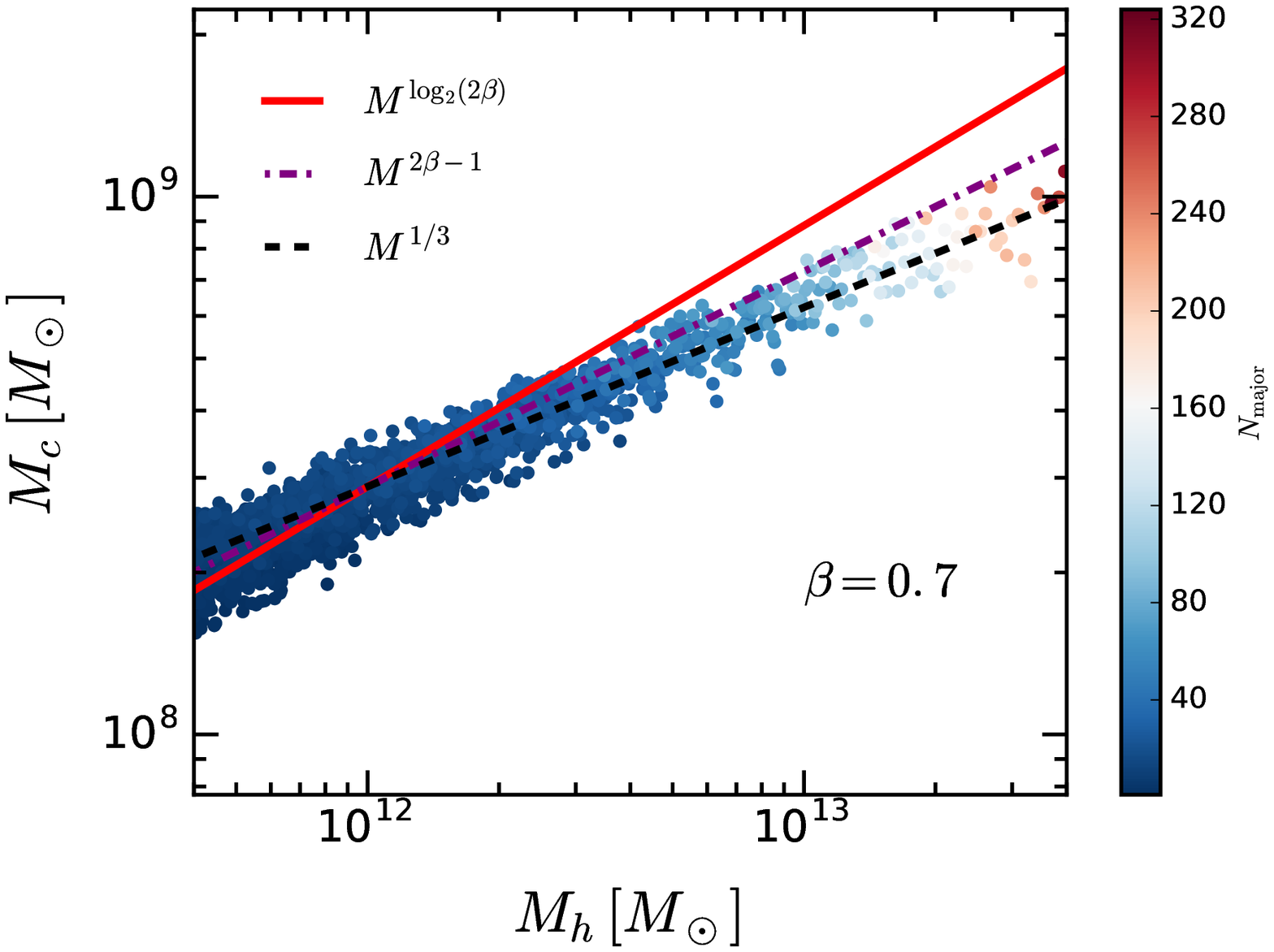}
\includegraphics[width=0.32\textwidth]{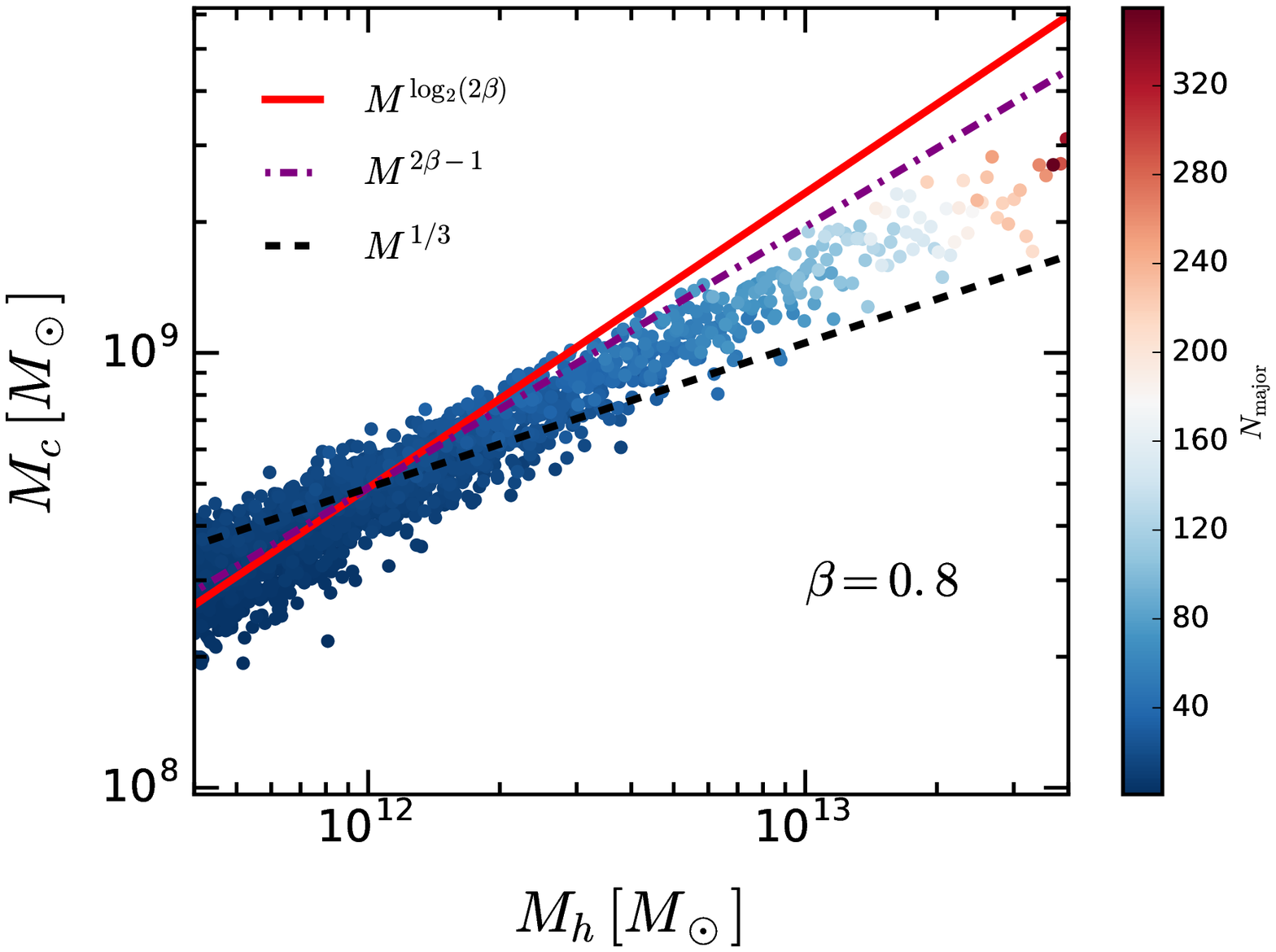}
\includegraphics[width=0.32\textwidth]{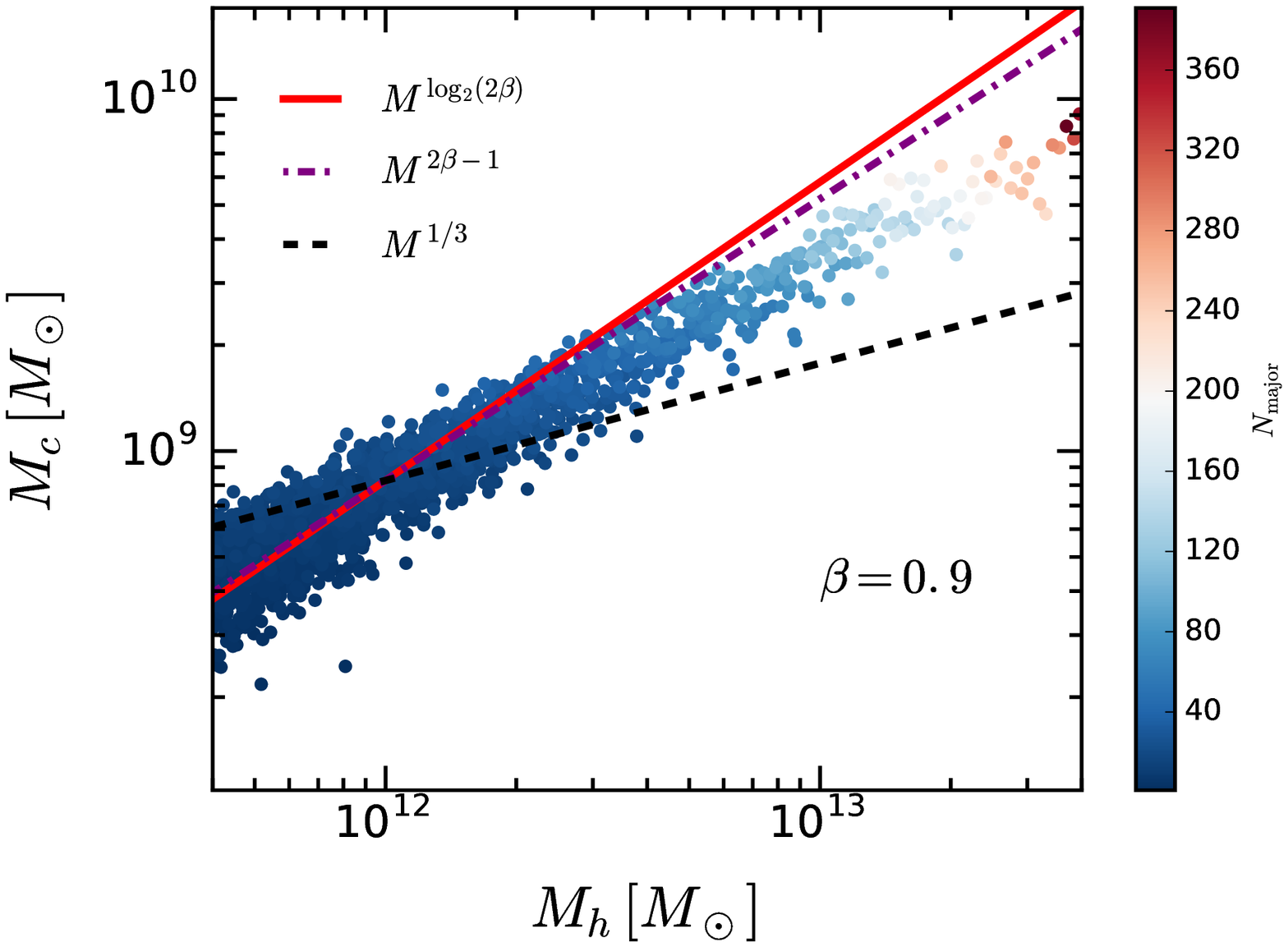}
\includegraphics[width=0.32\textwidth]{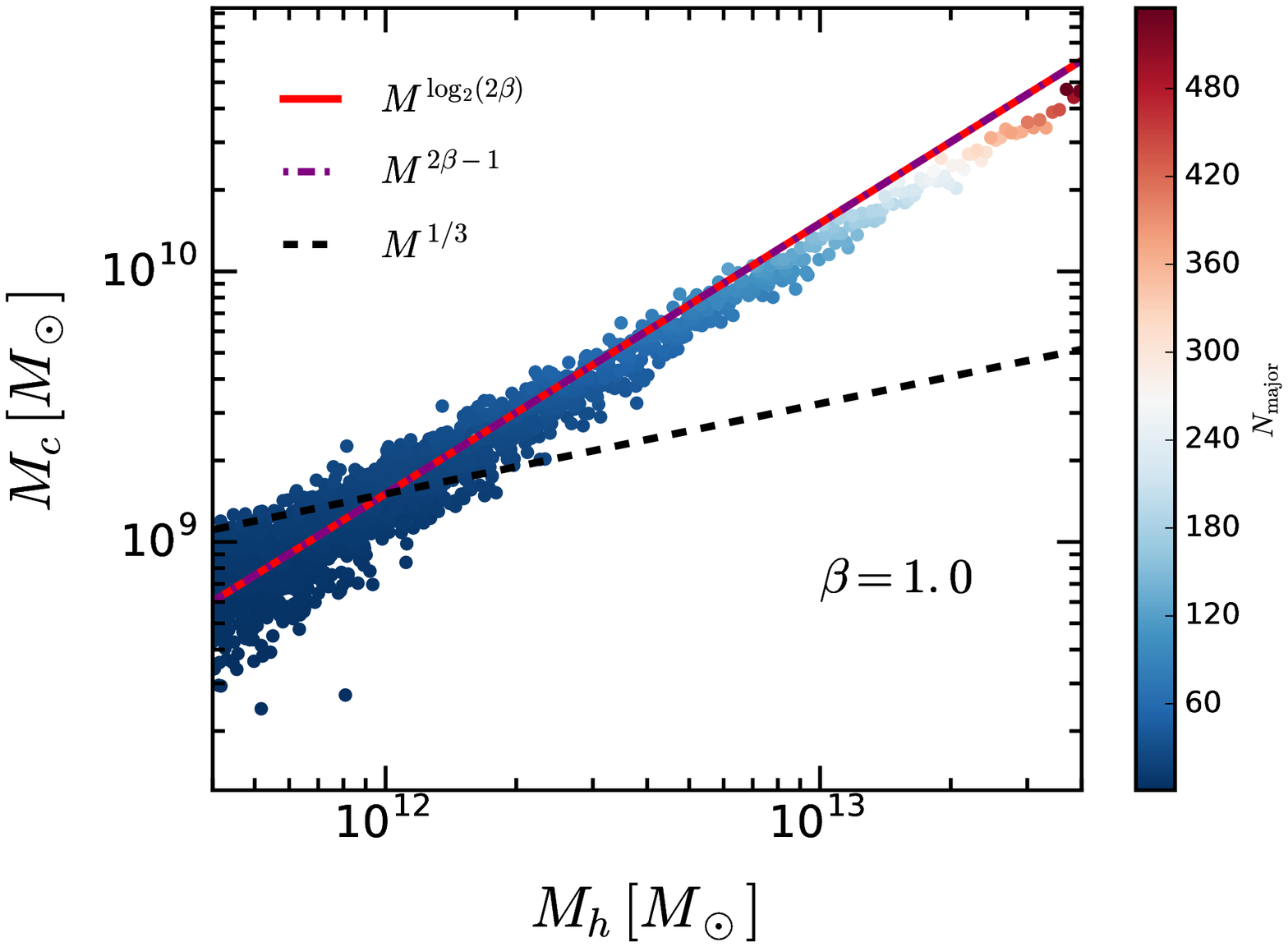}
\caption{The core-halo mass relation at $z=0$ for different $\beta$ compared with predictions of different models. The three lines are matched at $M_h=10^{12}M_{\odot}$.}
\label{fig:Mc_Mh_beta}
\end{figure*}

Figure~\ref{fig:Mc_Mh_beta} shows the core-halo mass relation at present
time for different $\beta$ and compares them with the predictions from
\cite{Schive2014b}, Eq.~(\ref{M_c_M_h}), and the linear parameterization
Eq.~(\ref{M_c_simple}).  
Despite the simplifications
in deriving Eq.~(\ref{M_c_M_h}), we find reasonable agreement for the core-halo
mass relation for different core mass loss fractions $(1-\beta)$. At
larger halo masses (implying more major mergers), the prediction of
our model Eq.~(\ref{M_c_M_h}) 
tends to overestimate the core masses. 
Equation~(\ref{M_c_simple})
gives a slightly better fit, 
implying that we can use it as a correction to Eq.~(\ref{M_c_M_h}).
For $\beta=0.7$ \cite{Schwabe:2016rze}, Eq.~(\ref{M_c_simple}) yields $M_c\propto
M_h^{0.4}$.
It is close to the relation $M_c\propto M_h^{1/3}$ and fits the cosmological simulations~\protect\cite{Schive2014a} equally well.

It should be noted that the merger trees constructed using the method described
in~\cite{Du:2016zcv} are not very accurate at redshifts
$z>3$.
In this work, we use the fitting formula for the mass-dependent critical overdensity~\cite{Marsh:2016vgj} instead of the one used in~\cite{Du:2016zcv} which is computed directly from the transfer function. We also improved the algorithm that is used to calculate the first crossing rate. We find the constructed merger trees to be more reliable at higher redshifts, i.e. the HMF calculated from merger trees matches the expectation from solving the excursion set problems.

In order to compare the core mass
predicted for FDM halos with observations, the prefactor $A$
in Eq.~(\ref{M_c_M_h}) is also
important. According to 
our results, we can replace 
$\log_2 (2\beta)$ in Eq.~(\ref{M_c_M_h}) with $2\beta-1$ to
give a better estimate of the core-halo mass relation.
If we further assume that at the beginning, i.e. prior to any mergers,
there were only pure solitons (instead of virialized halos produced by
mergers of solitons), the initial core mass is
$M_{\rm c,min}=\frac{1}{4}M_{\rm h,min}$ by
definition \cite{Schive2014a,Schive2014b}. Then we have 
\begin{equation}
M_c=\frac{1}{4}B\left(\frac{M_h}{M_{\rm h,min}}\right)^{2\beta-1}M_{\rm h,min},
\label{M_c_M_h3}
\end{equation}
where $B\equiv1/\{\alpha[1+b(\beta)]\}^{2\beta-1}$. The redshift dependence is implicitly contained in the function $B$. If
$\beta=2/3$, Eq.~(\ref{M_c_M_h3}) 
coincides exactly with the core-halo mass relation [Eq.~(6)] in \cite{Schive2014b}.

\section{Conclusions}\label{sec:conclusions}
By considering the merger history of dark matter halos in scenarios
with ultralight bosonic dark matter, we offer an alternative
explanation for the core-halo mass relation observed in cosmological simulations. 
We provide evidence for our model using stochastic merger trees and show that the
core-halo mass relation depends only on the mass loss fraction of cores
during binary mergers, $M_c\propto M_h^{2\beta-1}$.
We find that for $\beta=0.7$~\cite{Schwabe:2016rze}, 
this relation fits numerical data from cosmological simulations very well~\cite{Schive2014b}.

A similar approach may be employed to predict the statistical distribution of gravitationally bound substructures (axion stars or miniclusters) in scenarios with more massive axionlike particles or QCD axions. Instead of a single solitonic core, each dark matter halo hosts a large number of these objects and mergers take place both inside of individual halos and during halo mergers. Although a unique core-halo mass relation does not exist in this case, the universal mass gain for each substructure merger may still allow the construction of a stochastic model similar to ours. We will explore this possibility in future work.

\acknowledgements
We thank D.J.E. Marsh for helpful comments. X.D. acknowledges the China Scholarship Council (CSC) for financial support.

\bibliography{halo_core}

\end{document}